\RequirePackage{fix-cm}
\documentclass[smallextended]{svjour3}       
\smartqed  
\usepackage{graphicx}
\begin{document}

\title{ Cosmological expansion and local systems: a
Lema\^{i}tre-Tolman-Bondi model
}

\titlerunning{Cosmological expansion and local systems: a
Lema\^{i}tre-Tolman-Bondi model}        

\author{Ivana Bochicchio        \and
        Valerio Faraoni 
}


\institute{I. Bochicchio \at
              Dipartimento di Matematica,\\ Universit\'a degli Studi
di Salerno, \\ Via Ponte Don Melillo, 84084 Fisciano (SA), Italy \\
             \email{ibochicchio@unisa.it}           
           \and
           V. Faraoni \at
              Physics Department 
and {\em STAR} Research Cluster,\\  
Bishop's University, \\ a2600 College Street, Sherbrooke, \\
Qu\'ebec, Canada J1M 1Z7
}

\date{Received: date / Accepted: date}

\maketitle

\begin{abstract}
We propose a Lema\^{i}tre-Tolman-Bondi system mimicking a 
two-body 
system  to address the problem of the cosmological expansion 
versus local dynamics. This system is strongly bound 
but participates in the cosmic expansion and is exactly comoving 
with  the cosmic substratum. 

\end{abstract}

\section{Introduction}

The problem of the influence of the cosmological expansion on
local gravitationally bound systems was apparently first raised
by McVittie in 1933 \cite{McVittie33}, studied by Einstein and
Straus \cite{EinsteinStraus45,EinsteinStraus46}, and then
debated in dozens of papers stretching to our days (see the
recent review by Carrera and Giulini \cite{CarreraGiulini10}).

The problem is the following: for a local system such as a
planetary system, a binary stellar system, or a galaxy in the
expanding universe, does the cosmological expansion affect the
local dynamics of this system and, if so, in which way and,
numerically, to what extent? Many authors support the view that
the cosmic expansion affects only systems larger than a
certain spatial scale and there is no effect below that scale.
However, it has not been possible to assess what this scale is,
or  what determines it \cite{CarreraGiulini10,FaraoniJacques07}. It seems  that, assuming in
principle the Friedmann-Lema\^{i}tre-Robertson-Walker 
(FLRW)
metric to be valid down to small scales (possibly with some
modifications that describe local inhomogeneities
\cite{McVittie33,Lemaitre,Tolman,Bondi}),
all systems are subject to the effect of the cosmic expansion,
although this effect is numerically negligible for small systems
and stronger  for larger and larger objects, up to the scale of
galaxy clusters for
which it becomes significant. Even atoms have been considered as 
local
systems \cite{Bonnor,Price}, and a connection with the
well known anomaly observed in the Pioneer satellites has been
proposed \cite{Pioneer}, although there is no foundation for
attributing the Pioneer anomaly to the cosmic expansion
\cite{CarreraGiulini10}. Following the introduction of the
dark energy concept to explain the present acceleration of the
universe discovered with type Ia supernovae \cite{SN} and the
realization that this dark energy could potentially take the form
of phantom energy leading to a Big Rip singularity at a finite
future \cite{BigRip}, the effect of the cosmological
expansion on local systems as the Big Rip is approached seems to
go unquestioned even though the persistence of
cosmic effects on local systems in an  adiabatic approximation in
which the Hubble scale is much larger  than the
typical scales for local dynamics is often denied (see
\cite{FaraoniJacques07} for details).

Two lines of approach to the problem of cosmological expansion
versus local dynamics have been followed. The most common approach
studies a Newtonian gravitationally bound system, such as a
binary stellar system  embedded in an expanding FLRW universe and
computes the effect of the cosmic expansion as a perturbation of
the local dynamics. The result for a particle of polar
coordinates $ \left(r, \varphi \right)$ and angular momentum $L$
in the field of a mass $M$ embedded in a FLRW universe with scale
factor $a(t)$ is given by the equations of motion
\cite{CooperstockFaraoniVollick98,FaraoniJacques07,CarreraGiulini10}:
\begin{eqnarray}
\ddot{r} &=& -\frac{M}{r^2}   
+\frac{L^2}{r^2}+\frac{\ddot{a}}{a} \, r \,,\\
&&\nonumber\\
\dot{\varphi} &=& \frac{L}{r^2} \,.
\end{eqnarray}
The second approach, originally pursued with the McVittie
solution \cite{McVittie33} and the Einstein-Straus vacuole
\cite{EinsteinStraus45,EinsteinStraus46} used analytical
solutions of the Einstein equations to describe a local
(spherically symmetric) inhomogeneity embedded in a FLRW universe
(see \cite{Krasinski} for a review of inhomogeneous 
cosmological
models). The Lema\^{i}tre-Tolman-Bondi (LTB) class of 
solutions ({\em e.g.}, \cite{Ivanareview}) describes a 
spherical inhomogeneity embedded in a dust-dominated
(pressureless) FLRW universe. Here we  propose, in the LTB class
of solutions, a particularly clear example of a local system
composed  of two spherical shells with density merging smoothly
with the cosmic substratum, which is embedded in a FLRW universe.
These shells have zero angular momentum and are kept at finite
distance from each other by the cosmic expansion. Clearly, the
latter has a non-negligible effect on the local dynamics of
this system, independent of the strength of the local
interaction. In fact, the effect of the cosmological expansion on
the local two-shell system persists even in the limit in which
the ratio between local and cosmological densities is
arbitrarily large. For simplicity, we consider a spatially flat
FLRW background and we follow the notations of Ref. \cite{Wald}.
The metric signature is $-+++$ and units are used in which the
speed of light {\em in vacuo} $c$ and the gravitational constant 
$G$
assume the value unity.

\section{An exact two body system in a cosmological background}

The LTB line element is \cite{Lemaitre,Tolman,Bondi}
\begin{equation}\label{metric}
ds^2\,=\,-dt^{2}+\left[ R^{\prime }(t,r)\right]
^{2}dr^{2}+R^{2}\left( t,r \right) \, d\Omega ^{2}\ ,
\end{equation}
where $r$ is a comoving radius, $d\Omega ^{2}=d\vartheta ^{2}+\sin
^{2}\vartheta \,d\varphi ^{2}$ is the metric on the unit
$2$-sphere,
\begin{equation}\label{raggio}
R\left( t,r \right) = \left( r^{3/2}+\frac{3}{2} \,
\sqrt{m_{e}(r)}\,\,t\right) ^{2/3}
\end{equation}
is an areal radius,
\begin{equation}\label{massa}
m_{e}(r)=4\pi \int\limits_{0}^{r}dr'\,r'^2 \rho_0 (r') 
\,,
\end{equation}
is called the ``Euclidean mass",  and
$\rho _{0}(r)$ is the energy density on an initial hypersurface. 
The areal radius (\ref{raggio}) is obtained as 
solution of the classical Bondi equation 
\cite{Lemaitre,Tolman,Bondi,Ivanareview}
\begin{equation} \label{bondi}
\dot{R}^2(t,r)\,=\,\frac{ m_e(r)}{R(t,r)}\ .
\end{equation}   
A prime and a dot denote differentiation with 
respect to $r$ and $t$, respectively.

Let us consider, at the initial time, a density function that has 
two
equal peaks located at $r_1$ and $r_2$ (with $r_2>r_1$)
\begin{equation} \label{density}
\rho _{0}(r)=\rho _{c}+\rho _{*}\,\,\left[ \frac{1}{\left(
r-r_{1}\right) ^{2}+a^{2}}+\frac{1}{\left( r-r_{2}\right)
^{2}+a^{2}}\right]\ ,
\end{equation}
where $\rho _{c}$ is a positive constant denoting the 
density of a cosmological background far away from the 
density peaks, \emph{i.e.}, $\rho
_{0}(r)\,\approx\,\rho _c $ as $r \rightarrow +\infty$, and 
$\rho_{*}$ is proportional to the peak density with 
$\rho_*/a^2   >  \rho_c $.

\begin{figure}[t]
\centering
\includegraphics[width=0.7\textwidth]{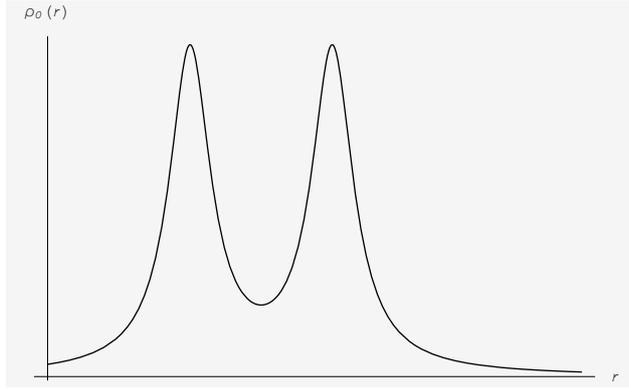}
\caption{The initial density $\rho_0(r)$ for the
parameters choice  $ \rho_{c} =\,1,  \rho_* = 2,  a^2 = 2,  
r_1=  8$,  and $ r_2=16 $ (in arbitrary units).} 
\label{figdensity}
\end{figure}
\begin{figure}[t]
\centering
\includegraphics[width=0.7\textwidth]{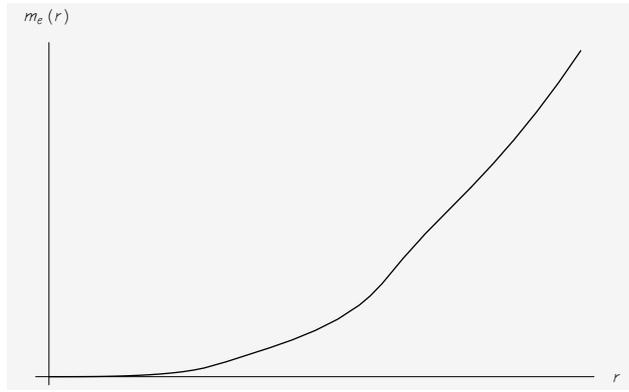}
\caption{The Euclidean mass $m_{e}(r)$. } \label{figmass}
\end{figure}

With the choice (\ref{density}) for the initial density
$\rho_0(r)$, the Euclidean mass can be calculated 
explicitly as (see Figs.~\ref{figdensity},~\ref{figmass})
\begin{eqnarray}  \label{mass}
m_{e}(r)&=& \frac{4\pi }{3}\rho _{c}r^{3}+8\pi \rho _{*}r
\nonumber\\
&&\nonumber\\
& + & 4 \pi \rho _{*}\left\{ \frac{\left(
r_{1}^{2}-a^{2}\right)}{a}\,\,\left[ \tan^{-1} \left(
\frac{r-r_1}{a}\right) + \tan^{-1} \left( 
\frac{r_1}{a}\right)
\right] + \right. \nonumber\\
&&\nonumber\\
& + & \left.  r_{1}\log \left[ \frac{\left( r-r_{1}\right)
^{2}+a^{2}}{r_{1}^{2}+a^{2}}\right] \right \} +
\nonumber\\
&&\nonumber\\
& + & 4 \pi \rho _{*} \left\{ \frac{\left(
r_{2}^{2}-a^{2}\right)}{a}\,\,\left[ \tan^{-1} \left(
\frac{r-r_{2}}{a}\right) +\tan^{-1} \left( 
\frac{r_2}{a}\right)
\right]\right.  \nonumber\\
&&\nonumber\\
& + & \left. r_{2}\log \left[ \frac{\left( r-r_{2}\right)
^{2}+a^{2}}{r_{2}^{2}+a^{2}}\right] \right \} \ .
\end{eqnarray}

This Euclidean mass is related to the 
Misner-Sharp-Hernandez mass and to
the Hawking-Hayward quasi-local mass. The 
Misner-Sharp-Hernandez mass
$M_{MSH}$ is defined, for a spherically
symmetric metric, by \cite{MisnerSharp64,HernandezMisner66}
\begin{equation}
1-\frac{2\,M_{MSH}}{R}=\nabla ^{\mu }R\nabla _{\mu
}R=1-R_{t}^{2}=1-\frac{\,m_{e}\left( r\right) }{R}
\end{equation}
from which it follows that
\begin{equation}
M_{MSH}=\frac{\,m_{e}\left( r\right) }{2}=2\pi
\int\limits_{0}^{r}ds\,\rho _{0}(s)s^{2} \,.
\end{equation}
The first equality is independent of the particular form of
the initial density distribution $\rho_0(r)$. Further, with
spherical symmetry, the Misner-Sharp-Hernandez mass 
coincides\footnote{See,  \emph{e.g.}, 
Ref.~\cite{CarreraGiulini10}.}  
with the Hawking-Hayward quasi-local mass \cite{Hawking68,Hayward86}.

\begin{figure}[t]
\centering
\includegraphics[width=0.7\textwidth]{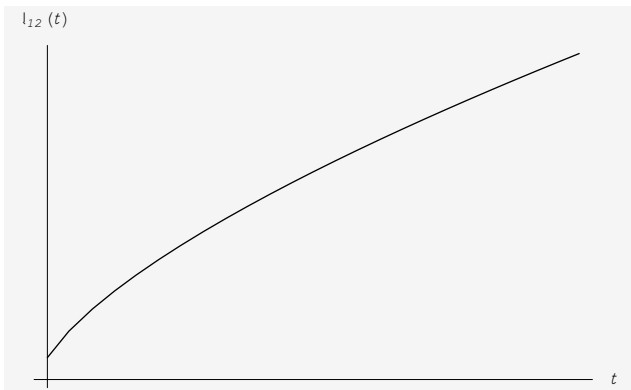}
\caption{The proper distance ${l}_{12}(t)$ as  a 
function of the comoving time. At $t=0$, $l_{12}$ coincides with  
the comoving coordinate distance $ r_2 - r_1 $.}
\label{proper-distance}
\end{figure}

With a slight abuse of terminology, the two density peaks at 
$r_1$ and
$r_2$ will be called ``shells". In actual fact, they are not
sharp shells, but continuous distributions of matter more or
less peaked according to the value of the parameter $a$ appearing
in (\ref{density}), and with maximum density controlled by 
the parameter $\rho _{*}$.
This two-``shell" system mimicks the binary system of test
particles chosen in many works to assess  the effect of the
cosmic expansion on local systems \cite{CarreraGiulini10}, but 
with two importance
differences. First, there is zero angular momentum,
no direction is preferred, and spherical symmetry holds because
instead of a single test particle orbiting around a centre 
of force and subject to the cosmic expansion as a 
perturbation of
this motion, we now have a spherical ``shell" surrounding a
smaller ``shell'' acting as a centre of force. Second, 
there is
no test particle here: the dust with density $\rho$ gravitates
and curves spacetime. Instead of treating the cosmic expansion
as a small perturbation of a weakly gravitating system, we have
an exact solution describing at once the spacetime and the
two-shells matter distribution, which merge smoothly with the
cosmological background, and this is the virtue of our LTB
example. No approximations are made and no adiabatic expansion in
term of local and Hubble time scales is necessary. The element of
proper radial distance is (setting $dt$, $d\vartheta$, and
$d\varphi$ equal to zero) $dl\,=\,R'dr$. Using the fact that
$dR\,=\,R_t dt+ R'dr$, we have $dl\,=\,dR$ for $dt\,=\,0$.
Therefore, the \emph{proper} radial distance between the two
density peaks on a constant time slice of the LTB spacetime is
\begin{eqnarray}
\emph{l}_{12}(t)&=&R_{2}-R_{1}\equiv R(t,\,r_{2})-R(t,\,r_{1})=
\nonumber\\
&&\nonumber\\
& = & \left( \,r_{2}^{3/2}+\frac{3}{2}\sqrt{ m_{e}\left(
r_{2}\right) } \, t\right) ^{2/3}-\left(
\,r_{1}^{3/2}+\frac{3}{2}\sqrt{ m_{e}\left( r_{1}\right)
} \, t\right) ^{2/3}\,.
\end{eqnarray}
This quantity can be regarded as the size of the 
two-shell system and
corresponds, roughly speaking,  to the radius of the binary system
considered in previous works. However, $l_{12}$ is now a proper
distance, not a coordinate distance (the use of misleading
coordinate systems has led to coordinate-based statements 
and has marred many investigations of this problem). At the time 
$t\,=\,0$ this
proper radial distance coincides with  the comoving coordinate
distance  $ \left( r_2-r_1 \right) $, but it departs from it for 
$t>0$. A plot of
$l_{12}$ is given in fig.~\ref{proper-distance}.

At large times $t \rightarrow +\infty$, it is
\begin{equation}
\l_{12}(t)\approx \left( \frac{3}{2}\right) ^{2/3}\left[ 
\left( \sqrt{m_{e}\left( r_{2}\right) }\right) 
^{2/3}-\left( \sqrt{m_{e}\left( r_{1}\right) }\right) 
^{2/3}\right] t^{2/3}\,, \label{asymptotics}
\end{equation}
\emph{i.e.}, the size $l_{12}$ of the two-shell system is
comoving and scales like the scale factor $a(t) \sim t^{2/3}$
of a dust-dominated FLRW universe, to which LTB spacetimes
reduce for large $r$. This result matches the one for the
Einstein-Straus vacuole \cite{EinsteinStraus45,EinsteinStraus46,Schucking54} in which a spherical central inhomogeneity
is surrounded by a spherical vacuum, which is matched to a
dust-dominated FLRW space. The proper radius of this matching
surface is perfectly comoving \cite{CarreraGiulini10}.

\begin{figure}[t]
\centering
\includegraphics[width=1\textwidth]{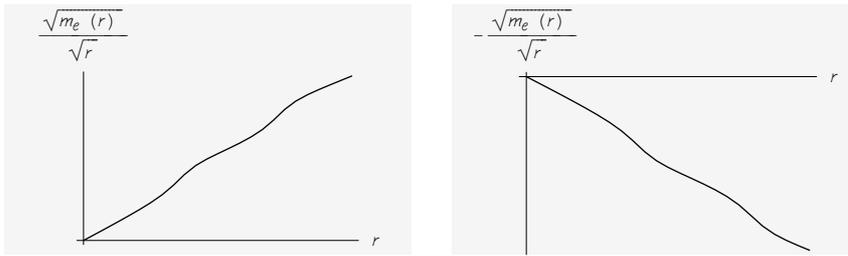}
\caption{The functions $\sqrt{m_e(r)/r}$ and $-\sqrt{m_e(r)/r}$ for  the 
choice~(\ref{density}) of the initial density and the same 
parameter values of fig.~\ref{figdensity}.} 
\label{function}
\end{figure}

We can now discuss the initial conditions in 
our model. The  initial radial separation  $ 
l_{12}(0)=R(0,  r_2)-R(0, r_1)=r_2-r_1 $ 
between the two ``shells'' can be taken 
arbitrarily. Following equation (\ref{bondi}),  
their relative (radial) velocity at time $t$ 
is
\begin{equation} 
\dot{l}_{12}(t) =   \dot{R}\left( t, r_2 \right)
-\dot{R}\left(t, r_1 \right)=\pm \Bigg(
\sqrt{  \frac{m_e(r_2) }{R(t, r_2)} } - 
\sqrt{  \frac{m_e(r_1)}{R(t, r_1)} }\Bigg)  
\,.\label{relativevelocity}
\end{equation}
At the initial time $t=0$, it is
\begin{equation} 
\dot{l}_{12}(0) = 
\pm \Bigg( \sqrt{  
\frac{m_e(r_2)}{r_2} } - 
\sqrt{  \frac{m_e(r_1)}{r_1} } \Bigg) \,.
\end{equation}
The functions $\sqrt{m_e(r)/r}$ and 
$-\sqrt{m_e(r)/r}$ are plotted in 
fig.~\ref{function}.  They are, respectively, 
always increasing and  always decreasing, 
therefore the sign of  the initial relative 
velocity  $ \dot{l}_{12}(0)$ of the two  
``shells'' is conserved. 
Assuming that the relative initial velocity of 
the ``shells'' is not sufficiently large that 
the shells collide and cause shell 
singularities (then the formalism 
outlined above remains valid), the relative 
velocity of the ``shells'' $\dot{l}_{12}$ given 
by eq.~(\ref{relativevelocity}) tends to zero 
because $R(t,r) \propto t^{2/3}$. The 
``shells'' then become comoving.

\section{Discussion}
Generally speaking, the Hubble flow washes out the effects 
of a local overdensity; this fact is well known in FLRW 
universes and it occurs also in LTB spacetimes. The 
following argument is a bit naive but nevertheless carries 
some weight: consider a function $f(r)$ and its radial 
gradient $ \frac{\partial f}{\partial R}$ with respect to 
the physical (areal) radius $R$ (the only spatial gradient 
allowed in the presence of spherical symmetry). It is 
straightforward to see that
\begin{equation}
\frac{\partial f}{\partial R}=\frac{ df/dr}{\partial 
R/\partial r}= 
\frac{
\left[ \sqrt{m_e(r)} \left( r^{3/2} +\frac{3}{2}\, \sqrt{ 
m_e(r)}\, t \right)^{1/3} \right] }{
\sqrt{ r m_e(r)} +2\pi r^2 \rho_0(r) t } \, \frac{df}{dr}
\end{equation}
and, therefore, $\partial f/\partial R \longrightarrow 0$ 
as $ t\rightarrow +\infty$. The evolution of our LTB model  
is in agreement with this general feature.

In our LTB model, in which the density contrast 
is smoothed out by the cosmic expansion, the energy density 
on an initial hypersurface
can be chosen at will and the expression (\ref{density})
describes two local concentrations  of matter (``shells")
merging smoothly with the cosmological background. These two
``shells'' play a role analogous to that of a binary system 
composed
of two test particles embedded in a FLRW background universe so
often considered in studies of the effects of the cosmological
expansion on local systems. We have chosen the peaks of the 
initial density distribution to be of equal height for 
simplicity but this assumption is not necessary: the ratio 
of their heights can be chosen at will, in the same way 
that the masses of a binary system  subject to the  
effect of the cosmological expansion can be chosen in any 
ratio.  The LTB example that we presented
has two clear advantages over the binary: first, it is a
fully gravitating configuration, not a system of test particles
and one does not need to resort to approximations in order to
embed this system into a cosmological background. This is an
exact solution of the Einstein equations. Second, due to the non
linearity of general relativity, when one wants to consider
cosmic expansion versus local dynamics, it is {\em a priori}
difficult or impossible to split a solution into a background 
plus an inhomogeneity (apart from perturbative regimes, which are 
nevertheless subject to the notorious gauge-dependence problems 
and require the use of gauge-invariant formalisms): this is 
exactly what the LTB metric  (or the McVittie metric 
\cite{McVittie33,FaraoniJacques07,CarreraGiulini10}) does for us.

It is reasonable to regard the proper radial distance
$l_{12}\,=\,R_2-R_1$ between the two ``shells" as the  physical
size of the local system. This quantity depends on time and it is
straightforward to see that this distance is comoving with the
cosmic dust of the FLRW background. In the LTB example proposed,
the local system is affected by the cosmic expansion, 
{\em regardless
of its spatial size}. One can take the limit $\rho_{*} 
\rightarrow 0 $ in which the system becomes a test fluid, 
or the opposite limit
$\rho_{*}/a^2 \gg \rho_c  $ in which the local system is 
strongly
gravitating. The initial distance $R_2-R_1\,=\,r_2-r_1$
between the two density peaks can be adjusted at will in
comparison with the Hubble radius $H^{-1}$ of the FLRW background.
Any way these parameters are varied, the result is always 
the 
same: the local system is stretched by
the cosmological expansion and participates in it. This fact shows
that, assuming the FLRW metric to describe the universe down to
small scales, it is not true that systems below a certain
spatial scale are unaffected while only  larger ones partake
into the expansion,  as suggested by many authors (see
\cite{FaraoniJacques07,CarreraGiulini10} for references). By
choosing the LTB metric as an example, we were bound to find
this result; indeed, it is built into the LTB metric 
itself.   One  may, therefore, question the validity
of our argument. The point is that a solution with these 
properties does exist, it  is a perfectly legitimate example,
and one of  the rare exact solutions of the Einstein equations in
which a clear answer to the puzzle of cosmic expansion versus
local dynamics can be obtained. The answer matches the results
obtained with the Einstein-Straus vacuole
\cite{EinsteinStraus45,EinsteinStraus46,Schucking54,CarreraGiulini10}, McVittie \cite{Nolan,CarreraGiulini10}, generalized McVittie solutions
\cite{FaraoniJacques07}, and LTB black holes 
\cite{Gaoetal2011}.
From the physical point of view, a relevant issue is
whether the two-shell system is stable, but this question can
also be posed for any LTB metric.

One potential problem with the LTB model proposed here 
consists of the possible presence of shell-crossing 
singularities in the $t<0$ region of this spacetime. 
Although these singularities are not mandatory in LTB 
models (conditions for their avoidance are given in 
Ref.~\cite{HellabyLake}), they are to be expected in 
general. The prevailing opinion about shell-crossing 
singularities is that they are an artifact of the dust 
equation of state chosen and should disappear when,  
more realistically,  some pressure is introduced (see, {\em 
e.g.},  Ref.~\cite{MTW}) although, to be honest, we do not 
know of a rigorous mathematical proof of this statement to 
date. Overall, we regard the possibility of shell-crossing 
singularities as a non-essential feature of LTB models and 
we choose to focus on the region of LTB spacetime which is 
free of these singularities.

Finally, we could make the two shells of our model 
infinitely thin and, in this case, they would have to be 
matched to their surroundings by using the Darmois-Israel 
junction conditions \cite{junction}. However, this 
construction is not necessary and by continuity of the 
property presented here,  one expects that a system 
composed of  two zero-thickness shells  will also be comoving. 
The case of a single shell matched  through the
junction conditions to an expanding FLRW universe, and
hosting a wormhole, was considered in
Refs.~\cite{FaraoniIsrael05,BochicchioFaraoni10} with the result
that, even if the shell initially has a non-vanishing velocity 
relative to the cosmic substratum, it eventually becomes 
comoving with it.
The  answer to the question of 
cosmic 
expansion versus local dynamics seems to  be that  
expansion wins and local systems go with the (Hubble) flow.

\begin{acknowledgements}

We thank a referee for useful remarks. I. B. thanks the 
Fonds Qu\'eb\'{e}cois de la Recherche sur la
Nature et les Technologies (FRQNT) for financial
support and Bishop's University for its hospitality. V. F. 
is supported by the Natural Sciences and Engineering 
Research Council of Canada (NSERC) and by Bishop's 
University.
\end{acknowledgements}




\end{document}